\documentclass[conference,a4paper]{IEEEtran}
\usepackage{xcolor}
\usepackage{balance}
\usepackage{cite}
\usepackage{multirow}
\usepackage[pdftex]{graphicx}
\usepackage{amsmath,amssymb,amsfonts}
\usepackage{textcomp}
\usepackage[caption=false,font=footnotesize]{subfig}
\usepackage[nolist]{acronym}
\usepackage[capitalize]{cleveref}
\usepackage{nicefrac}
\usepackage{graphicx}  
\crefname{table}{Tab.}{Tabs.}
\Crefname{table}{Tab.}{Tabs.}
\crefformat{equation}{(#2#1#3)}

\newcommand\norm[1]{\left\lVert#1\right\rVert}

\newcommand{\Estimate}[2]{\widetilde{\mathbf{#1}}^{\left( {\rm #2} \right)} }

\newcommand{\HatX}[2]{\widehat{\mathbf{#1}}^{\left( {\rm #2} \right)} }

\newcommand{\BarX}[2]{\overline{\mathbf{#1}}^{\left( {\rm #2} \right)} }

\newcommand{\NoLine}[2]{\mathbf{#1}^{\left( {\rm #2} \right)} }
\newcommand{\TRx}[3]{#1_{\rm #2}^{\left( {\rm #3} \right)} }

\newcommand{\NormalBrackets}[1]{ \left( #1 \right) }

\begin{document}
\begin{acronym}
    \acro{3GPP}[3GPP]{3rd Generation Partnership Project}
    \acro{5G}[5G]{fifth generation}

    \acro{ADC}[ADC]{analog-to-digital converter}
    \acro{AMC}[AMC]{adaptive modulation and coding}    
    \acro{AoA}[AoA]{angle-of-arrival}
    \acro{AoD}[AoD]{angle-of-departure}
    \acro{AQNM}[AQNM]{additive quantization noise model}
    \acro{AR}[AR]{augmented reality}
    \acro{AWG}[AWG]{arbitrary waveform generator}
    \acro{AWGN}[AWGN]{additive white Gaussian noise}
 
    
    \acro{CDF}[CDF]{cumulative distribution function}
    \acro{CIR}[CIR]{channel impulse response}
    \acro{CS}[CS]{compressed sensing}
    \acro{CSI}[CSI]{channel state information}    
    \acro{CSI-RS}[CSI-RS]{channel state information reference signal}
    \acro{CTF}[CTF]{channel transfer function}
    \acro{CW}[CW]{continuous wave}
    
    \acro{DAC}[DAC]{digital-to-analog converter}
    \acro{DFT}[DFT]{discrete Fourier transform}
    \acro{DR}[DR]{dynamic range}
    \acro{DSD}[DSD]{Doppler power spectral density}

    \acro{EE}[EE]{energy efficiency}
    \acro{EIRP}[EIRP]{equivalent isotropically radiated power}
    
    
    
    \acro{HPBW}[HPBW]{half-power beamwidth}

    \acro{IDFT}[IDFT]{inverse discrete Fourier transform}
    \acro{IF}[IF]{intermediate frequency}
    \acro{IL}[IL]{insertion loss}    
    
    \acro{LMMSE}[LMMSE]{linear minimum mean square error}
    \acro{LNA}[LNA]{low noise amplifier}
    \acro{LO}[LO]{local oscillator}
    \acro{LOS}[LOS]{line-of-sight}
    \acro{LS}[LS]{least-squares}
    \acro{LSF}[LSF]{local scattering function}
    
    \acro{MIMO}[MIMO]{multiple-input multiple-output}
    \acro{MISO}[MISO]{multiple-input single-output}
    \acro{MMW}[mmWave]{millimeter wave}
    \acro{MoM}[MoM]{method of moments}
    \acro{MRC}[MRC]{maximal ratio combining}
    \acro{MSE}[MSE]{mean squared error}

    \acro{NLOS}[NLOS]{non-line-of-sight}    
    \acro{NR}[NR]{new radio}
    
    \acro{OFDM}[OFDM]{orthogonal frequency-division multiplexing}
    \acro{OLOS}[OLOS]{obstructed line-of-sight}
    \acro{OMP}[OMP]{orthogonal matching pursuit}
    \acro{OOBA-LOS}[OOBA-LOS]{out-of-band aided LOS}    
    \acro{OOBA-MRC}[OOBA-MRC]{out-of-band aided maximal ratio combining}

    \acro{PA}{power amplifier}
    \acro{PCB}[PCB]{printed circuit board}    
    \acro{PDP}[PDP]{power delay profile}
    \acro{PHY}[PHY]{physical layer}

    \acro{QAM}[QAM]{quadrature amplitude modulation}
    
    \acro{RF}[RF]{radio frequency}
    \acro{RMS}[RMS]{root-mean-square}
    \acro{Rx-to-Tx}[Rx-to-Tx]{receiver-to-transmitter}
    
    \acro{SE}[SE]{spectral efficiency}
    \acro{SINR}[SINR]{signal-to-interference-and-noise ratio}
    \acro{SMD}[SMD]{surface mount device}
    \acro{SNR}[SNR]{signal-to-noise ratio}
    \acro{SVD}[SVD]{singular value decomposition}
    
    \acro{TDD}[TDD]{time-division duplex}
    \acro{Tx-to-Rx}[Tx-to-Rx]{transmitter-to-receiver}

    \acro{ULA}[ULA]{uniform linear array}
    \acro{UPA}[UPA]{uniform planar array}

    \acro{V2I}[V2I]{vehicle-to-infrastructure}    
    \acro{V2X}[V2X]{vehicle-to-everything}
    \acro{VAA}[VAA]{virtual antenna array}    
    \acro{VNA}[VNA]{vector network analyzer}
    \acro{VR}[VR]{virtual reality}

    \acro{XR}[XR]{extended reality}

\end{acronym}

\title{Exploiting Out-of-Band Information for Millimeter-Wave MIMO Channel Estimation: Performance in Static and Dynamic Scenarios}

\author{\IEEEauthorblockN{
Faruk Pasic\IEEEauthorrefmark{1},
Mariam Mussbah\IEEEauthorrefmark{1}\IEEEauthorrefmark{2},
Stefan Schwarz\IEEEauthorrefmark{1},
Markus Rupp\IEEEauthorrefmark{1} and
Christoph F. Mecklenbräuker\IEEEauthorrefmark{1}
}%

\IEEEauthorblockA{\IEEEauthorrefmark{1}
Institute of Telecommunications, TU Wien, Vienna, Austria}
\IEEEauthorblockA{\IEEEauthorrefmark{2}
Christian Doppler Laboratory for Digital Twin assisted AI for sustainable Radio Access Networks}
\IEEEauthorblockA{faruk.pasic@tuwien.ac.at}
}

\IEEEoverridecommandlockouts 

\makeatletter
\def\thanks#1{\protected@xdef\@thanks{\@thanks
        \protect\footnotetext{#1}}}
\makeatother

\maketitle

\begin{abstract}
To support the high data rates for latency-critical applications, future wireless systems will employ fully digital beamforming \ac{MIMO} architectures at \ac{MMW} frequencies.
Moreover, \ac{MMW} \ac{MIMO} deployments will coexist with conventional sub-6\,GHz \ac{MIMO} systems, creating opportunities to exploit out-of-band sub-6\,GHz information to enhance channel estimation at \ac{MMW} frequencies.
In this work, we analyze the pilot-aided channel estimation performance of \ac{MMW} \ac{MIMO} systems under various pilot configurations in both static and dynamic environments. 
We evaluate the system performance in terms of \ac{SE} for line-of-sight and non-line-of-sight propagation conditions. 
Simulation results show that incorporating out-of-band sub-6\,GHz information yields notable \ac{SE} gains in both static and dynamic scenarios.
\end{abstract}
\vskip0.5\baselineskip

\acresetall

\section{Introduction}
\Ac{MMW} communication systems are a promising solution for accommodating the rapidly growing demand for high data rates, enabled by the availability of large bandwidths in the \ac{MMW} spectrum~\cite{Molisch2025, Hammoud2025_TVT, Pasic2023_mag}.
To compensate for the severe path loss inherent at these frequencies, \ac{MMW} systems rely on large-scale antenna arrays in conjunction with \ac{MIMO} techniques~\cite{Heath2016}.
Fully digital beamforming architectures are essential to fully exploit the spatial multiplexing and beamforming capabilities of \ac{MIMO} systems~\cite{Yang2018, Pasic2025_3}.
Unlike analog and hybrid architectures, digital beamforming enables simultaneous multi-directional channel estimation within a single time interval, thereby significantly reducing link establishment overhead and supporting latency-critical applications~\cite{Liu2020}.

A major challenge in the realization of digital \ac{MMW} \ac{MIMO} systems is reliable link establishment, which depends on accurate channel estimation~\cite{Heath2016}.
At \ac{MMW} frequencies, channel estimation is particularly demanding due to the low pre-beamforming \ac{SNR}.
In contrast, sub-6\,GHz systems experience significantly lower propagation losses, resulting in a higher pre-beamforming \ac{SNR}~\cite{Hofer2025, Pasic2025_OJCOMS}.
As \ac{MMW} systems are increasingly deployed alongside sub-6\,GHz systems~\cite{Shafi2020}, this reliable out-of-band information can be leveraged to assist and enhance \ac{MMW} channel estimation.
Several recent studies have investigated the use of out-of-band information for \ac{MMW} channel estimation~\cite{Pasic2024, Pasic2025_TCOM}, proposing pilot-aided methods that exploit the \ac{LOS} channel component acquired with the support of the sub-6\,GHz band.

The performance of digital \ac{MMW} beamforming systems has been studied in static environments~\cite{Dutta2017_2, Sohrabi2017, Cebeci2023, Tataria2020}, as well as in dynamic scenarios~\cite{Khorsandmanesh2025, Okuyama2020, Tian2024}. 
However, existing studies on digital \ac{MMW} beamforming systems that exploit out-of-band information have largely focused on static environments~\cite{Ali2016, Pasic2025_TCOM, Pasic2025_INFOCOM}, while dynamic scenarios remain largely unexplored.
Moreover, to the best of the author's knowledge, the impact of different pilot configurations on the performance of digital \ac{MMW} \ac{MIMO} systems in dynamic scenarios has not yet been studied.

\textbf{Contribution:}
In this paper, we evaluate the performance of digital beamforming \ac{MMW} \ac{MIMO} systems in both static and dynamic environments.
We consider both conventional channel estimation and out-of-band aided channel estimation as proposed in~\cite{Pasic2025_TCOM}.
Our analysis considers various pilot configurations, a wide range of receiver velocities, different \ac{SNR} levels and both \ac{LOS} and \ac{NLOS} propagation conditions characterized by the Rician $K$-factor.
The performance of the considered methods is assessed through simulations in terms of \ac{SE}.

\textbf{Organization:}
\cref{sec:system_model} describes the system model.
\cref{sec:methods} outlines the considered channel estimation methods.
A simulation-based performance evaluation is presented in~\cref{sec:comparison}. 
Finally, \cref{sec:conclusion} concludes the paper.

\textbf{Notation:} 
The superscript $\left( \cdot \right) ^{\left( \rm b \right)}$ indicates quantities that depend on the operating frequency band, where ${\rm b} \in \{ {\rm s}, {\rm m} \}$, with ${\rm s}$ denoting the sub-6\,GHz and ${\rm m}$ the \ac{MMW} band.
Scalars are represented by $x$, vectors by bold lowercase letters ${\mathbf x}$ and matrices by bold uppercase letters ${\mathbf X}$.
For a matrix ${\mathbf X}$, $\mathbf{X}_{i, :}$ and $\mathbf{X}_{:, j}$ denote its $i$-th row and $j$-th column, respectively.
The transpose and Hermitian transpose operators are denoted by $\left( \cdot \right) ^{\rm T}$ and $\left( \cdot \right) ^{\rm H}$, respectively.
The Euclidean norm is written as $\lVert \cdot \rVert$, while the Frobenius norm is denoted by $\lVert \cdot \rVert_F$.

\section{System Model} \label{sec:system_model}
We consider a point-to-point multi-band \ac{MIMO} communication system in which sub-6\,GHz and \ac{MMW} systems operate simultaneously under far-field propagation conditions.
The transmitter employs $\TRx{M}{Tx}{s}$ antenna elements at sub-6\,GHz and $\TRx{M}{Tx}{m}$ antenna elements at \ac{MMW}. 
Correspondingly, the receiver is equipped with $\TRx{M}{Rx}{s}$ sub-6\,GHz and $\TRx{M}{Rx}{m}$ \ac{MMW} antenna elements.
Both frequency bands utilize \acp{ULA} composed of dipole antennas, which are approximated as isotropic radiators.
The sub-6\,GHz and \ac{MMW} arrays are assumed to be spatially co-located and aligned, such that the \ac{LOS} path is characterized by the same \ac{AoD} $\vartheta$ and \ac{AoA} $\varphi$ in both bands~\cite{Ma2021}.
The spacing between adjacent antenna elements in each array is chosen as $\TRx{\Delta d}{}{b} = \,$0.5$\,\TRx{\lambda}{}{b}$, where $\TRx{\lambda}{}{b}$ denotes the wavelength of the respective system.
We further assume perfect time and frequency synchronization across all transmitting and receiving antennas.
The transmitter and receiver are equipped with one \ac{RF} chain per antenna, enabling fully digital beamforming in both the sub-6\,GHz and \ac{MMW} systems.
The system follows a \ac{TDD} operation, thereby ensuring reciprocal channel responses.

We consider an \ac{OFDM} transmission system employing $\TRx{N}{}{b}$ subcarriers and $\TRx{K}{}{b}$ \ac{OFDM} time-symbols.
For each \ac{OFDM} time-symbol, a block of $\TRx{N}{}{b}$ \ac{QAM} symbols is assigned to $\TRx{N}{}{b}$ distinct subcarriers.
The \ac{OFDM} system converts a broadband frequency-selective channel into narrowband frequency-flat channels with the help of a \ac{DFT} and application of a cyclic prefix~\cite{Cimini1985}.
The cyclic prefix length is assumed to exceed the maximum excess delay of the multipath channel, thereby eliminating inter-symbol interference. 
Furthermore, inter-carrier interference, such as that caused by Doppler shifts or carrier frequency offsets, is considered negligible~\cite{Moose1994}.

\subsection{Channel Model} \label{subsec:channel_model}
For a given \ac{OFDM} subcarrier $n$ and time-symbol $k$, the propagation channel is represented by an $\TRx{M}{Rx}{b} \times \TRx{M}{Tx}{b}$ dimensional complex-valued channel matrix $\NoLine{H}{b} [n, k]$, based on the equivalent complex baseband model of the \ac{OFDM} system.
These matrices characterize the small-scale fading behavior of the wireless channel incorporating the \ac{OFDM} processing at the transmitter and receiver.
In particular, $\NoLine{H}{b} [n, k]$ corresponds to a sampled version of the underlying \ac{CTF}, following the formulation in~\cite{Cho2010}.
In what follows, if the description of a quantity is independent of the subcarrier index $n$ and/or the time-symbol index $k$, the respective index is omitted for notational simplicity.

We adopt a time-varying frequency-selective channel model with Rician fading, modeled as~\cite{molisch2012wireless}
\begin{equation}
    \begin{split}        
        \NoLine{H}{b} [n, k] & = 
          \sqrt{\TRx{\eta}{}{b}} 
         \sqrt{\frac{\TRx{\kappa}{}{b}}{1+\TRx{\kappa}{}{b}}} 
         \underbrace{\NoLine{H}{b}_{\rm fs} [k]}_{\rm free-space} \\
        & +
        \sqrt{\TRx{\eta}{}{b}}
        \sqrt{\frac{1}{ 1+\TRx{\kappa}{}{b}} } 
        \underbrace{\NoLine{H}{b}_{\rm sp} [n, k]}_{\rm stochastic\, part},
    \end{split}       
    \label{eq:channel_model}
\end{equation}
where $\TRx{\eta}{}{b}$ accounts for large-scale attenuation effects, including path loss and shadowing, while $\TRx{\kappa}{}{b}$ denotes the Rician $K$-factor associated with frequency band ${\rm b} \in \{ {\rm s}, {\rm m} \}$.
To reflect the different propagation characteristics at \ac{MMW} and sub-6\,GHz frequencies, the \ac{MMW} $K$-factor is assumed to scale with that of the sub-6\,GHz band according to $\TRx{\kappa}{}{m}=c_{\kappa} \TRx{\kappa}{}{s}$, where $c_{\kappa}$ denotes the scaling factor.
The deterministic \ac{LOS} channel component corresponds to a free-space propagation model and is defined by
\begin{equation}
    \NoLine{H}{b}_{\rm fs} [k] =
    e^{j \TRx{\chi}{}{b}}
    e^{j\frac{2 \pi \TRx{\nu}{}{b} k}{\bigtriangleup \TRx{f}{}{b}}}
    \TRx{\mathbf{a}}{Rx}{b} \NormalBrackets{ \varphi }
    \NormalBrackets{\TRx{\mathbf{a}}{Tx}{b} \NormalBrackets{ \vartheta }}^{\rm H}
    \label{eq:free_space_component}
\end{equation}
where
\begin{equation}
    \TRx{\mathbf{a}}{Tx}{b} \NormalBrackets{ \vartheta } =  
    \begin{bmatrix}
        1 & \cdots  & 
        e^{-j 2 \pi \left( \TRx{M}{Tx}{b} - 1 \right) 
        \frac{ \TRx{\Delta d}{}{b}}{ \TRx{\lambda}{}{b}} 
        \sin \NormalBrackets{ \vartheta } } 
        \end{bmatrix}^{\rm T}
        \label{eq:AoD_steering_vector}
\end{equation}
and
\begin{equation}
    \TRx{\mathbf{a}}{Rx}{b} \NormalBrackets{ \varphi } =  
    \begin{bmatrix}
        1 & \cdots  & 
        e^{-j 2 \pi \NormalBrackets{ \TRx{M}{Rx}{b} - 1 } 
        \frac{\TRx{\Delta d}{}{b}}{\TRx{\lambda}{}{b}} 
        \sin \NormalBrackets{ \varphi } }     
        \end{bmatrix}^{\rm T}
\end{equation}
denote the array steering vectors associated with the \ac{AoD} $\vartheta$
and \ac{AoA} $\varphi$, respectively.
In~\cref{eq:free_space_component}, the term $\TRx{\chi}{}{b} ~\sim~\mathcal{U}~\left( - \pi, \pi \right)$ models the random phase offset at the first (reference) antenna element, $\TRx{\nu}{}{b}$ denotes the Doppler shift and $\bigtriangleup \TRx{f}{}{b}$ represents the subcarrier spacing.
The stochastic channel matrix $\NoLine{H}{b}_{\rm sp} [n, k]$ is generated according to the \ac{3GPP} channel model~\cite{3gpp.38.901}.
Given the large difference between the sub-6\,GHz and \ac{MMW} frequencies, the stochastic channels $\NoLine{H}{s}_{\rm sp} [n, k]$ and $\NoLine{H}{m}_{\rm sp} [n, k]$ are assumed to be statistically independent.
Furthermore, a channel model following Jakes sum-of-sinusoids approach~\cite{Jakes1994} is employed to determine the temporal evolution of the wireless channel~\cite{Zemen2005}.

\subsection{Link Establishment} \label{subsec:link_establishment}
The establishment of a communication link consists of a training phase, followed by a data transmission phase.
During the training phase, the wireless channel is estimated independently at sub-6\,GHz and \ac{MMW} frequency bands.
The channel estimates from both bands are then utilized jointly for data transmission at the \ac{MMW} band.

\subsubsection{Training Phase} \label{subsubsec:training_phase}

\begin{figure}[t]
    \centering
    {\includegraphics[width=\columnwidth]{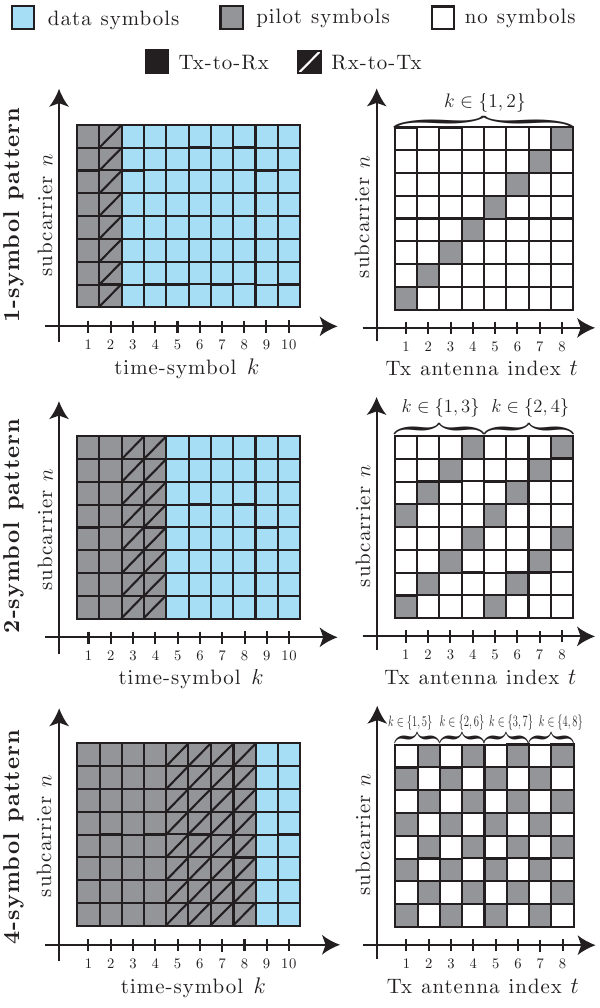}}
    \caption{Pilot symbols at each antenna occupy a certain number of subcarriers such that they do not overlap in the frequency domain. 
    The 1-symbol pattern (top) confines all pilot transmissions to a single OFDM time-symbol, the 2-symbol pattern (middle) expands the pilot allocation over two OFDM time-symbols, while the 4-symbol pattern (bottom) spans four OFDM time-symbols per transmission direction.}
    \label{fig:pilot_allocation}
\end{figure}

In the training phase, known pilot symbols are transmitted from each antenna element and are assumed to be perfectly known at the receiver.
The pilot symbols $\TRx{\boldsymbol{\phi}}{}{b} [n, k] \in \mathbb{C}^{\TRx{M}{Tx}{b} \times 1} $ generated from the \ac{QAM} constellation are distributed across $\TRx{N}{}{b}$ subcarriers such that each transmit antenna is assigned a distinct subset of non-overlapping subcarriers.

For the pilot allocation, we consider three pilot patterns similar to those defined by the \ac{3GPP} for the \ac{CSI-RS}~\cite{3gpp.38.211}.
These pilot patterns are designed to facilitate efficient and scalable channel estimation for different antenna configurations and deployment scenarios.
Note that the pilot patterns considered in this work are provided as illustrative examples and are not fully compliant with the \ac{3GPP} standard.
These pilot patterns are intended for potential use in future systems and may be adapted based on the specific configuration and deployment scenario.
In this work, we refer to them colloquially as the 1-symbol, 2-symbol, and 4-symbol pilot patterns, based on the number $\TRx{K}{P}{b}$ of \ac{OFDM} time-symbols they span.
Note that the system operates using the \ac{TDD} protocol, which requires the pilot symbols to be allocated for both the \ac{Tx-to-Rx} and \ac{Rx-to-Tx} directions.
Since we typically use the same number of transmit and receive antennas in this paper, we set the number of time-symbols $\TRx{K}{P}{b}$ to be identical for both the \ac{Tx-to-Rx} and \ac{Rx-to-Tx} directions.
This constraint can, however, be relaxed when the numbers of transmit and receive antennas differ.
An example pilot allocation for a system with $M_{\rm Tx}^{\left( \rm b \right)} = 8$ transmit antennas is illustrated in~\cref{fig:pilot_allocation}.
The pilot allocation at the $t$-th transmit antenna is given by
\begin{equation}
    \mathrm{\phi}_t^{\NormalBrackets{\rm b}} [n,k] =
    \begin{cases}
        \TRx{\mathrm{\phi}}{}{b} [n,k], \quad \text{if } (n, k, t) \in \TRx{\mathcal{P}}{}{b}
        \\
        0, \quad \quad \quad \quad \, \, \text{else}
    \end{cases}
    \label{eq:pilot_allocation_2_symbol}
\end{equation}
where the set of pilot positions $\TRx{\mathcal{P}}{}{b}$ is defined by
\begin{equation}
    \begin{split}
    \TRx{\mathcal{P}}{}{b} = 
    \left\{ \right. & \left. (n, k, t) : \right. \\  
    & \left. n  \in \left\{ t_m, t_m + \frac{\TRx{M}{Tx}{b}}{\TRx{K}{P}{b}}, \ldots, \TRx{N}{}{b} - \frac{\TRx{M}{Tx}{b}}{\TRx{K}{P}{b}} + t_m \right\}, 
    \right. \\
    & \left.
    k \in \{1, 2, \ldots, 2 \TRx{K}{P}{b} \},  \quad
    t \in \left\{ 1, 2, \ldots, \TRx{M}{Tx}{b} \right\},
    \right. \\
    & \left. 
    t_m  = \NormalBrackets{t - 1 \bmod \frac{\TRx{M}{Tx}{b}}{\TRx{K}{P}{b}} } + 1
    \right\}.
    \end{split}
\end{equation} 
and $t \in \{ 1, \ldots, M_{\rm Tx}^{\left( \rm b \right)} \}$ denotes the transmit antenna index.

The 1-symbol pilot pattern $\NormalBrackets{ \TRx{K}{P}{b} = 1 }$ confines all pilot transmissions to a single \ac{OFDM} time-symbol per transmission direction. 
This enables fully parallel pilot transmission and offers minimal time-domain overhead, making it attractive for latency-critical applications.
However, the limited number of pilot resource elements restricts its applicability to systems with few transmit antennas and may result in reduced estimation accuracy in frequency-selective channels due to sparse pilot density in the frequency domain.

The 2-symbol pilot pattern $\NormalBrackets{ \TRx{K}{P}{b} = 2 }$ expands the pilot allocation over two \ac{OFDM} time-symbols per transmission direction.
This extension increases the available pilot resources, supporting a larger number of antennas and improving frequency-domain resolution.
This denser pilot placement leads to more reliable channel estimation in frequency-selective environments, at the expense of higher time-domain overhead compared to the 1-symbol pattern.

In the 4-symbol pilot pattern $\NormalBrackets{ \TRx{K}{P}{b} = 4 }$, pilots span four \ac{OFDM} time-symbols, providing high pilot density and robust channel estimation, particularly for large-scale \ac{MIMO} systems.
The improved accuracy comes with a significant increase in overhead, which reduces resources available for data transmission and increases latency.

The input-output relationship of the training phase is given by
\begin{equation} 
    \NoLine{y}{b} [n, k] = 
    \NoLine{H}{b} [n, k] 
    \TRx{\boldsymbol{\phi}}{}{b} [n, k] +
    \NoLine{w}{b} [n, k],
    \label{eq:training_input_output}
\end{equation}
where the received signal is denoted by $\NoLine{y}{b} [n, k] \in \mathbb{C}^{\TRx{M}{Rx}{b} \times 1} $ and the \ac{AWGN} with the power of $\sigma_{\TRx{w}{}{b}}^2$ is denoted by $\NoLine{w}{b} [n, k] \sim \mathcal{CN}\NormalBrackets{0,\sigma_{\TRx{w}{}{b}}^2 \mathbf{I}_{\TRx{M}{Rx}{b}}}$.    
At the receiver, \ac{LS} channel estimates are first obtained at the pilot subcarriers.
The channel response at the remaining subcarriers is then obtained via linear interpolation, yielding the estimated channel matrix $\Estimate{H}{b} [n] \in \mathbb{C}^{\TRx{M}{Rx}{b} \times \TRx{M}{Tx}{b}}$.
Since the channel estimation spans over $\TRx{K}{P}{m}$ symbols to obtain a full \ac{MIMO} channel estimate, we consider the estimated quantities as time-invariant and therefore omit the time-index $k$.
However, it is important to note that the actual channel $\NoLine{H}{b} [n, k]$ may still vary over time.

Since channel estimation is performed at the receiver, the corresponding \ac{CSI} has to be available also at the transmitter to enable optimal beamforming.
Owing to the use of a \ac{TDD} scheme, the transmitter acquires the \ac{CSI} by leveraging channel reciprocity in both frequency bands.
Specifically, the procedure begins with the transmission of training symbols from the transmitter to the receiver (\ac{Tx-to-Rx}), allowing the receiver to acquire the \ac{CSI}. 
This is followed by the transmission of training symbols in the reverse direction (\ac{Rx-to-Tx}), enabling the transmitter to acquire the \ac{CSI}. 

\subsubsection{Data Transmission} \label{subsec:data_transmission}
During the data transmission phase, only the \ac{MMW} link is considered. 
Data symbols are transmitted over $\TRx{N}{}{m}$ subcarriers and $\TRx{K}{D}{m} = \TRx{K}{}{m} - \TRx{K}{P}{m}$ \ac{OFDM} time-symbols.
The channel estimates obtained in the training phase are further processed using the methods which will be detailed in~\cref{sec:methods}.
The resulting \ac{MMW} channel estimate $\BarX{H}{m} [n]$ is then employed to design the precoding and combining matrices.
To maximize \ac{SE}, \ac{SVD} is adopted.
The compact-form \ac{SVD} of $\BarX{H}{m} [n]$ is given by
\begin{equation}
    \BarX{H}{m} [n] = 
    \BarX{Q}{m} [n] \,
    \BarX{\Sigma}{m} [n]
    \NormalBrackets{\BarX{F}{m} [n]}^{\rm H},
    \label{eq:svd}
\end{equation}
where $\BarX{Q}{m} [n] \in \mathbb{C}^{\TRx{M}{Rx}{m} \times {\ell_{\rm max}}}$ contains the left singular vectors and serves as the combining matrix, $\BarX{F}{m} [n] \in \mathbb{C}^{\TRx{M}{Tx}{m} \times {\ell_{\rm max}}}$ comprises the right singular vectors and defines the precoder, $\BarX{\Sigma}{m} [n]$ denotes the diagonal matrix of singular values ${\rm diag} \left( \TRx{\overline{\sigma}}{(1)}{m} [n], \ldots, \overline{\sigma}_{(\ell_{\rm max})}^{\left( \rm m \right)} [n] \right)$ and $\ell_{\rm max} = {\rm min}  \NormalBrackets{ \TRx{M}{Rx}{m}, \TRx{M}{Tx}{m} }$ denotes the maximum number of spatial streams.
The power loading matrix is defined by $\BarX{P}{m} [n] = {\rm diag} \NormalBrackets{ \overline{p}_{(1)}^{\left( \rm m \right)} [n], \ldots , \overline{p}_{(\ell_{\rm max})}^{\left( \rm m \right)} [n] }$ and is optimized via the water-filling algorithm to maximize the achievable rate under a total transmit power constraint
\begin{equation}
    \norm{ 
    \BarX{F}{m} [n]
    \NormalBrackets{\BarX{P}{m} [n]}^{1/2}
    }_F^2 = 
    \TRx{P}{\rm T}{m},
    \label{eq:power_normalization}
\end{equation}
where $\TRx{P}{\rm T}{m}$ denotes the total transmit power.
With this notation, the input-output relationship for the data transmission phase is then given by
\begin{equation} 
    \begin{split}
    \NoLine{y}{m} [n, k] & =
    \NormalBrackets{ \BarX{Q}{m} [n] }^{\rm H}
    \NoLine{H}{m} [n, k] \,
    \BarX{x}{m} [n, k] \\
    & + 
    \NormalBrackets{ \BarX{Q}{m} [n] }^{\rm H}
    \NoLine{w}{m} [n, k],
    \end{split}
    \label{eq:data_input_output}
\end{equation}
where the precoded transmit signal is denoted by $\BarX{x}{m} [n, k] = \BarX{F}{m} [n] \NormalBrackets{ \BarX{P}{m} [n]}^{1/2} \NoLine{x}{m} [n, k]$, the unprecoded transmit signal is denoted by $\NoLine{x}{m} [n, k] \in \mathbb{C}^{\ell_{\rm max} \times 1}$, the received signal is denoted by $\NoLine{y}{m} [n, k] \in \mathbb{C}^{\ell_{\rm max} \times 1} $ and the \ac{AWGN} added at the \ac{MMW} receiver is denoted by $\NoLine{w}{m} [n, k] \sim \mathcal{CN} \NormalBrackets{ 0, \sigma_{\TRx{w}{}{m}}^2 \mathbf{I}_{\TRx{M}{Rx}{m}}}$.
Since $\BarX{Q}{m} [n]$ is semi-unitary, $ \NormalBrackets{ \BarX{Q}{m} [n] }^{\rm H} \NoLine{w}{m} [n, k]$ has the same statistical distribution as $\NoLine{w}{m} [n, k]$.

\section{Channel Estimation Methods} \label{sec:methods}
This section outlines the \ac{MMW} channel estimation schemes considered in this work.

\subsection{Conventional} \label{subsec:conventional}
The conventional approach uses only the in-band \ac{MMW} channel estimate $\Estimate{H}{m} [n]$, obtained via \ac{LS} estimation at pilot subcarriers and subsequent linear interpolation, as described in~\cref{subsubsec:training_phase}.
The resulting \ac{MMW} channel estimate is given by
\begin{equation}
    \BarX{H}{m} [n] = \Estimate{H}{m} [n]. 
    \label{eq:conventional_eq}
\end{equation}
Note that this approach does not leverage any additional side information or advanced signal processing techniques.

\subsection{Out-of-band Aided MRC (OOBA-MRC)} \label{subsec:ooba_mrc}
This method applies \ac{OOBA-MRC} to optimally combine the in-band channel estimate $\Estimate{H}{m} [n]$ (detailed in~\cref{subsubsec:training_phase}) with the out-of-band aided channel estimate $\HatX{H}{m} [n]$ (introduced in~\cite{Pasic2025_TCOM}). 
The combined \ac{MMW} channel estimate is given by
\begin{equation}
    \BarX{H}{m} [n] =
    \widehat{w} \,
    \HatX{H}{m} [n]
    + (1-\widehat{w})
    \Estimate{H}{m} [n]
   \label{eq:mrc}
\end{equation} 
where 
\begin{equation}
    \widehat{w} \approx
    \frac{ \TRx{M}{Tx}{m} \TRx{M}{Rx}{m} \sigma_{w^{\left( \rm m \right)}}^2}
    { \frac{\TRx{M}{Tx}{m} \TRx{M}{Rx}{m} }{1 + 
    \TRx{\widetilde{\kappa}}{}{s} }
    + \NormalBrackets{ 1 + \TRx{M}{Tx}{m} \TRx{M}{Rx}{m} } \sigma_{w^{\left( \rm m \right)}}^2 }.
    \label{eq:optimal_w}
\end{equation}
denotes the approximated optimal combining factor, whose derivation is detailed in~\cite{Pasic2025_TCOM}.
In~\cref{eq:optimal_w}, $\TRx{\widetilde{\kappa}}{}{s}$ denotes the Rician $K$-factor in the sub-6\,GHz band, obtained via the method of moments~\cite{Greenstein1999}.
This method is designed to adapt to dynamic wireless channel conditions with regard to the Rician $K$-factor and the noise variance $\sigma_{w^{\left( \rm m \right)}}^2$.

\begin{table}[t]
    \centering
    \footnotesize
    \caption{Simulation Parameters} 
    \label{tab:simulationParams}
    \begin{tabular}{rcc}
        \hline
        \textbf{Parameter}                          & \multicolumn{2}{c}{\textbf{Value}} \\ \hline
        Frequency Band                              & sub-6 GHz         & mmWave         \\
        Carrier Frequency $f_{\rm c}$               & 2.55\,GHz         & 25.5\,GHz           \\
        Wavelength $\lambda$                        & 11.76\,cm         & 1.176\,cm          \\
        Bandwidth $B$                               & 20.16\,MHz        & 403.2\,MHz            \\
        Subcarrier Spacing $\bigtriangleup f$       & 60\,kHz           & 120\,kHz             \\
        Number of Data Symbols $K_{\rm D}$          & N/A                 & 7           \\        
        Transmit Power $P_{\rm T}$                  & 30\,dBm           & 30\,dBm  \\ 
        Antenna Configuration $\NormalBrackets{ M_{\rm Rx} \times M_{\rm Tx}}$    & 8$\times$8               & 8$\times$8 \\         
        RMS Delay Spread $\sigma_{\tau}$            & 1148\,ns          & 841\,ns     \\
        Number of Realizations $L_r$                & 1000              & 1000             \\  \hline
    \end{tabular}
\end{table}

\begin{figure*}[t]
    \centering
    \subfloat[]{
        \includegraphics[width=0.45\textwidth]{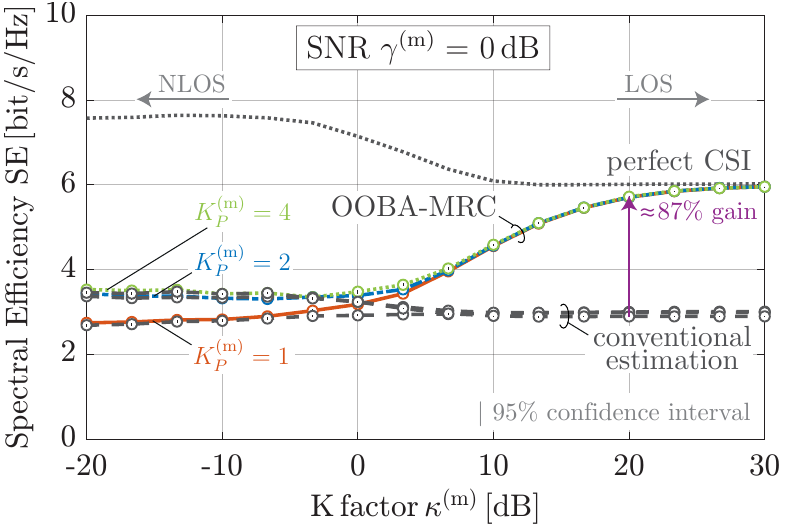} 
        \label{fig:SE_vs_K_factor_Pilots_a}
    }
    \hfill 
    \subfloat[]{
        \includegraphics[width=0.45\textwidth]{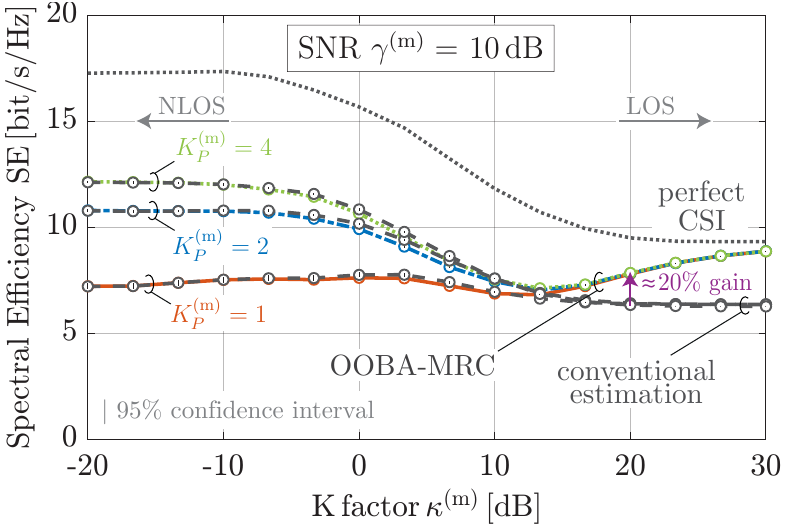} 
        \label{fig:SE_vs_K_factor_Pilots_b}
    }
    \caption{At a lower \ac{SNR} of 0\,dB, increasing the number of pilot symbols does not yield a significant improvement in \ac{SE} (left), whereas at a higher \ac{SNR} of 10\,dB, the performance gains become more pronounced  (right).}
    \label{fig:SE_vs_K_factor_Pilots}
\end{figure*}

\section{Simulation Results} \label{sec:comparison}
In this section, we evaluate the performance of the channel estimation methods introduced in~\cref{sec:methods} and the impact of different pilot schemes using Monte-Carlo simulations. 
We consider the achievable \ac{SE} as the main performance metric for \ac{MIMO} systems.
Throughout the analysis, we assume equal power allocation across all subcarriers.
For the frequency-selective channel model, we employ the \ac{3GPP} model for the urban macro \ac{LOS} scenario with a configurable $K$-factor, as specified in~\cite{3gpp.38.901}.
Additionally, we assume that the \ac{AoD} and \ac{AoA} are mutually independent and uniformly distributed within the range $\left[ -90^\circ, 90^\circ \right]$.
Given the close similarity between sub-6\,GHz and \ac{MMW} $K$-factors observed in the measured scenario reported in~\cite{Pasic2025}, we set the scaling factor $c_{\kappa}=1$ ($c_{\kappa}=0\,$dB), thereby defining the \ac{MMW} $K$-factor as $\TRx{\kappa}{}{m}= \TRx{\kappa}{}{s}$.
The corresponding simulation parameters are summarized in~\cref{tab:simulationParams}, which also includes the parameters for the sub-6\,GHz \ac{MIMO} system required for the \ac{OOBA-MRC} method.
In addition to the methods described in~\cref{sec:methods}, we also include the performance for the case of perfect \ac{CSI} $\BarX{H}{m} [n] = \NoLine{H}{m} [n]$ to establish an upper performance bound.

The achievable \ac{SE} in bits$/$s$/$Hz averaged over $\TRx{N}{}{m}$ subcarriers and $\TRx{K}{D}{m}$ time-symbols is given by
\begin{equation} 
    \mathrm{SE}= 
    \frac{1}{\TRx{N}{}{m} \TRx{K}{D}{m}} 
    \sum_{n=1}^{\TRx{N}{}{m}} 
    \sum_{k=\TRx{K}{P}{m} + 1}^{\TRx{K}{}{m}}     
    \sum\limits_{\mu=1}^{\ell_{\rm max}} 
    \log_2 \left( 1 + \mathrm{SINR}_{\rm \mu} [n, k] \right)
    \label{eq:se}
\end{equation}
with the effective \ac{SINR} for the stream ${\rm \mu}$ denoted by~\cite{Kammoun2014}
\begin{equation} 
    \mathrm{SINR}_{\rm \mu} [n, k] = \frac{ 
    \left| \overline{\mathrm{G}}_{\rm \mu,\mu}^{\left( {\rm m} \right)} [n, k] \right|^2 }
    { \sum\limits_{\substack{\nu=1 \\ \nu \neq \mu}}^{\ell_{\rm max}} 
    \left| \overline{\mathrm{G}}_{\rm \mu, \nu}^{\left( {\rm m} \right)} [n, k] \right|^2
    + \sigma^2  
    \norm{ \overline{\mathrm{Q}}_{\rm :,\mu}^{\left( {\rm m} \right)} [n, k] }^2 }.    
    \label{eq:sinr}
\end{equation}
In~\cref{eq:sinr}, the element $\overline{\mathrm{G}}_{\rm \mu,\nu}^{\left( {\rm m} \right)} [n, k]$, with $\mu, \nu \in \{ 1, \ldots, \ell_{\rm max} \} $, denotes an entry of the channel gain matrix $\overline{\mathbf{G}}^{\left( {\rm m} \right)} [n, k] \in \mathbb{C}^{\ell_{\rm max} \times \ell_{\rm max}}$ for the $n$-th subcarrier and $k$-th time-symbol.
The matrix is defined as
\begin{equation}
    \BarX{G}{m} [n, k] =
    \NormalBrackets{ \BarX{Q}{m} [n] } ^ {\rm H} 
    \NoLine{H}{m} [n, k] \,
    \BarX{F}{m} [n] 
    \NormalBrackets{\BarX{P}{m} [n] }^{1/2}.
    \label{eq:chgain}
\end{equation}
The overall error variance is given by $\sigma^2 = \sigma_{\rm \mu}^2 [n, k] + \sigma_{w^{\left( \rm m \right)}}^2$, where $ \sigma_{\rm \mu}^2 [n, k]$ corresponds to the diagonal elements of the estimation error covariance matrix $\BarX{C}{m}_{\varepsilon} [n, k] \in \mathbb{C}^{\ell_{\rm max} \times \ell_{\rm max}}$.
The covariance matrix is computed as
\begin{equation}
    \BarX{C}{m}_{\varepsilon} [n, k] = 
    \frac{1}{\ell_{\rm max}}
    \TRx{\overline{\boldsymbol{\varepsilon}}}{}{m} [n, k]
    \NormalBrackets{ \TRx{\overline{\boldsymbol{\varepsilon}}}{}{m} [n, k] }^{\rm H},
    \label{eq:error_covariance}
\end{equation}
where the estimation error matrix is defined by
\begin{equation}
    \TRx{\overline{\boldsymbol{\varepsilon}}}{}{m} [n, k] = 
    \NoLine{G}{m} [n, k] - \BarX{G}{m} [n, k].
    \label{eq:chgain_error}
\end{equation}
In~\cref{eq:chgain_error}, $\TRx{\overline{\boldsymbol{\varepsilon}}}{}{m} [n, k]$ quantifies the error between the channel gain matrix obtained using the estimated precoder and combiner, $\BarX{G}{m} [n, k]$, and that obtained with the perfectly designed precoder and combiner, $\NoLine{G}{m} [n, k]$.

\subsection{Performance in Static Scenarios} \label{sect.impact_pilot_symbols}
In this subsection, we investigate the impact of different pilot schemes (see~\cref{subsubsec:training_phase}) on the considered channel estimation methods as a function of the $K$-factor in static scenarios.
The simulation results in terms of achievable \ac{SE} are shown  in~\cref{fig:SE_vs_K_factor_Pilots}.

\begin{figure*}[t]
    \centering
    \subfloat[]{
        \includegraphics[width=0.47\textwidth]{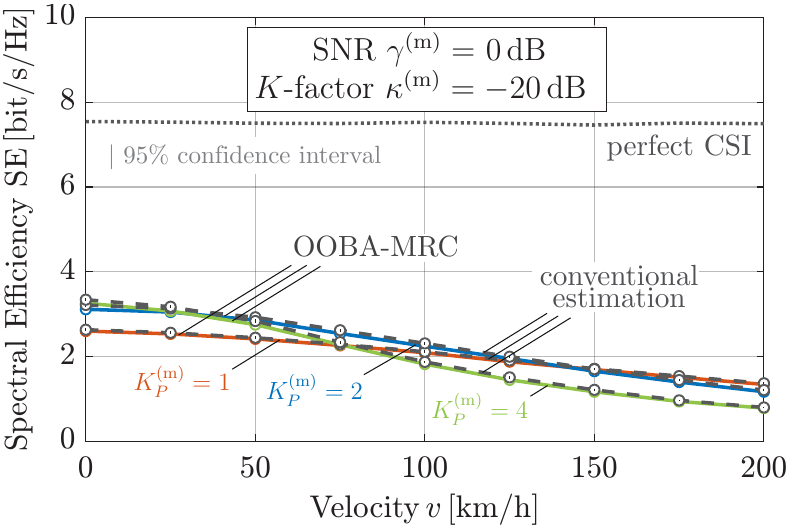} 
        \label{fig:SE_vs_Velocity_a}
    }
    \hfill 
    \subfloat[]{
        \includegraphics[width=0.47\textwidth]{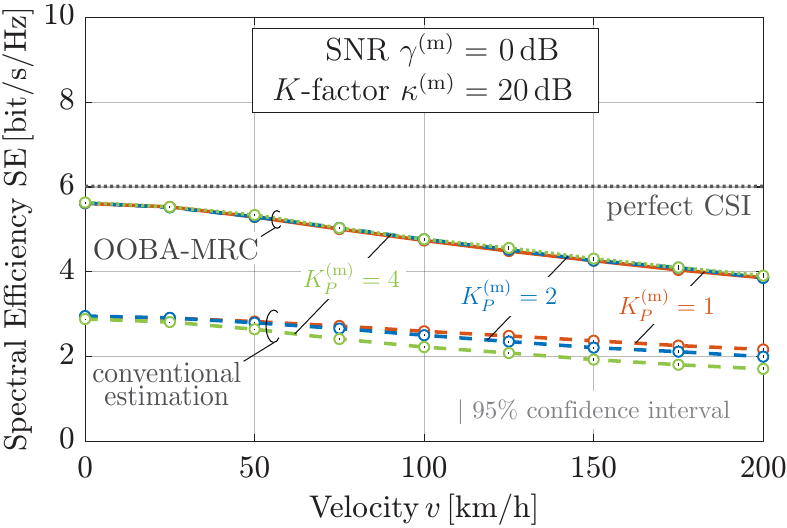} 
        \label{fig:SE_vs_Velocity_b}
    }
    \caption{At a low $K$-factor of $-$20\,dB, the \ac{OOBA-MRC} method provides no gain in achievable \ac{SE} compared to the conventional method (left) and at a high $K$-factor, its relative gain increases with the number of pilot symbols (right).}
    \label{fig:SE_vs_Velocity}
\end{figure*}

The \ac{OOBA-MRC} method achieves equal or better performance than the conventional method, regardless of the pilot pattern.
For $K$-factors below 0\,dB, increasing the number of pilot symbols generally improves the \ac{SE}, since more pilots yield more accurate channel estimates.
As the $K$-factor increases beyond 0\,dB, the performance obtained with different pilot patterns converges, because the channel becomes increasingly dominated by the \ac{LOS} component and adding more pilot symbols in the frequency domain does not provide additional benefit.
For an \ac{SNR} of 0\,dB and low $K$-factor, the \ac{SE} improves from 2.7\,bit/s/Hz (1-symbol pattern) to 3.4\,bit/s/Hz (2-symbol pattern), and only slightly further to 3.5\,bit/s/Hz (4-symbol pattern). 
For an \ac{SNR} of 10\,dB, the gains from additional pilot symbols are more pronounced: the \ac{SE} increases from 7\,bit/s/Hz (1-symbol pattern) to 11\,bit/s/Hz (2-symbol pattern) and to 12.5\,bit/s/Hz (4-symbol pattern).

\subsection{Performance in Dynamic Scenarios} \label{sect.performance_velocity}

In this subsection, we analyze the performance of the considered estimation methods under time-varying channel conditions as a function of the receiver velocity.
Simulation results for an \ac{MMW} \ac{SNR} of 0\,dB  in terms of achievable \ac{SE} are shown in~\cref{fig:SE_vs_Velocity}.
Both \ac{NLOS} and \ac{LOS} propagation conditions are considered, characterized by $K$-factors of $-$20\,dB (see~\cref{fig:SE_vs_Velocity_a}) and 20\,dB (see~\cref{fig:SE_vs_Velocity_b}), respectively.

Two counteracting effects are of particular interest here. 
Increasing the number of pilot symbols enhances the estimation accuracy by providing more reliable channel information.
However, increasing the number of pilot symbols causes channel aging, where the estimates lose relevance as the channel changes over time.
Thus, it is important to evaluate the resulting trade-off between estimation accuracy and channel aging.
As expected, the achievable \ac{SE} generally decreases with increasing velocity, due to faster channel variations, regardless of the $K$-factor, estimation method or pilot pattern employed.

In the \ac{NLOS} scenario with a $K$-factor of $-$20\,dB, the \ac{OOBA-MRC} method primarily relies on the in-band channel estimate, resulting in performance that closely matches that of the conventional method, regardless of the pilot pattern.
For low-velocities (up to approximately 50\,km/h), the \ac{SE} increases with the number of pilot symbols $\TRx{K}{P}{m}$, as the enhanced estimation accuracy dominates.
In the time-invariant case, the 1-symbol pilot pattern yields an \ac{SE} of 2.7\,bit/s/Hz, the 2-symbol pattern increases the \ac{SE} to 3.4\,bit/s/Hz and the 4-symbol pattern reaches 3.5\,bit/s/Hz.
With increasing velocity, however, channel aging becomes dominant, leading to a faster \ac{SE} degradation for larger pilot patterns.
At 200\,km/h, the 1-symbol pattern achieves the highest \ac{SE} of 1.3\,bit/s/Hz, while the 2-symbol pattern slightly decreases to 1.2\,bit/s/Hz and the 4-symbol pattern performs worst at 0.8\,bit/s/Hz. 
The 1-symbol pattern begins to outperform the 4-symbol pattern at approximately 75\,km/h and surpasses the 2-symbol pattern from about 150\,km/h onward.

In the \ac{LOS} scenario with a $K$-factor of 20\,dB, the \ac{OOBA-MRC} method outperforms the conventional method by dominantly exploiting the out-of-band aided \ac{LOS} channel estimate. 
Its performance remains essentially independent of the number of pilot symbols.
In contrast, the conventional method is affected by channel aging, causing performance to deteriorate as the number of pilot symbols increases. 
When the receiver velocity reaches 200\,km/h, the \ac{SE} drops from 2.9\,bit/s/Hz in the time-invariant case to 2.2\,bit/s/Hz for the 1-symbol pattern, 2\,bit/s/Hz for the 2-symbol pattern and 1.7\,bit/s/Hz for the 4-symbol pattern.
For the \ac{OOBA-MRC} method, varying the number of pilot symbols has no noticeable impact, with the \ac{SE} decreasing from 5.6\,bit/s/Hz in the static case to 3.9\,bit/s/Hz at 200\,km/h.

\section{Conclusion} \label{sec:conclusion}
In this paper, we investigate the channel estimation performance of digital beamforming \ac{MMW} \ac{MIMO} systems and the impact of different pilot schemes in both static and dynamic scenarios. 
Simulation results confirm the effectiveness of the \ac{OOBA-MRC} method compared to the conventional method.
The \ac{SNR} and Rician $K$-factor affect the performance of the \ac{OOBA-MRC} method to a great extent.

In static scenarios, increasing the number of pilot symbols improves the quality of the channel estimates.
For low $K$-factors, where the channel is highly frequency-selective, this results in an overall increase in \ac{SE}.
However, for high $K$-factors where the channel is predominantly flat, increasing the number of pilot symbols does not improve the \ac{SE} performance.

In dynamic scenarios, channel aging becomes a dominant factor.
As the number of pilot symbols increases, the channel may change significantly during the estimation interval, causing older estimates to lose relevance. 
Consequently, for low $K$-factors, lower velocities benefit from additional pilots, leading to increased \ac{SE}, while higher velocities experience a decline in \ac{SE} as more pilots are used. 
For high $K$-factors, the conventional method exhibits decreasing \ac{SE} with an increasing number of pilots, whereas \ac{OOBA-MRC} remains robust and maintains high performance even at high mobility.

\section*{Acknowledgment}
The work of M.~Mussbah has been funded by the Christian Doppler Laboratory for Digital Twin assisted AI for sustainable Radio Access Networks, Institute of Telecommunications, TU Wien. 
The financial support by the Austrian Federal Ministry for Labour and Economy and the National Foundation for Research, Technology and Development and the Christian Doppler Research Association is gratefully acknowledged.

\bibliography{references}
\bibliographystyle{IEEEtran}

\end{document}